\documentstyle[12pt]{article}
\begin{document}
\renewcommand{\Re}{{\rm Re \,}}
\renewcommand{\Im}{{\rm Im \,}}
\newcommand{\Tr}{{\rm Tr \,}}
\newcommand{\beq}{\begin{equation}}
\newcommand{\eeq}[1]{\label{#1}\end{equation}}
\newcommand{\bea}{\begin{eqnarray}}
\newcommand{\eea}[1]{\label{#1}\end{eqnarray}}
\newcommand{\dirac}{/\!\!\!\partial}
\newcommand{\Dirac}{/\!\!\!\!D}
\begin{titlepage}
\begin{center}
\hfill hep-th/9512180,  NYU-TH-95/12/01, CERN-TH/95-346
\vskip .4in
{\large\bf Spontaneous Breaking of N=2 to N=1 in Rigid and Local
Supersymmetric Theories}
\end{center}
\vskip .4in
\begin{center}
{\large Sergio Ferrara}$^a$,
{\large Luciano Girardello}$^b$ and
{\large Massimo Porrati}$^c$,
\vskip .1in
$a$ Theory Division, CERN, CH-1211 Geneva 23, Switzerland
\vskip .05in
$b$ Theory Division, CERN, CH-1211 Geneva 23, Switzerland~\footnotemark
\footnotetext{On leave from
Dipartimento di Fisica, Universit\`a di Milano, via Celoria 16,
20133 Milano, Italy.}
\vskip .05in
$c$ Department of Physics, NYU, 4 Washington Pl.,
New York NY 10003, USA~\footnotemark
\footnotetext{On leave from INFN, Sez. di Pisa, 56100 Pisa, Italy.}
\end{center}
\vskip .4in
\begin{center} {\bf ABSTRACT} \end{center}
\begin{quotation}
\noindent
We analyze the relation between rigid and local supersymmetric N=2 field
theories, when half of the supersymmetries are spontaneously broken. In
particular, we show that the recently found partial supersymmety breaking
induced by electric and magnetic Fayet-Iliopoulos terms in rigid theories
can be obtained by a suitable flat limit of previously constructed N=2
supergravity models with partial super-Higgs in the
observable sector.
\end{quotation}
\vfill
\end{titlepage}
\section*{Introduction}
Recent work on D-brane physics has renewed interest in the analysis of
supersymmetry breaking in N=2 supergravity models that may describe the
low-energy effective actions of string theories that take into account
non-perturbative phenomena, such as the role of R-R charged states in
conifold transitions, R-R Fayet-Iliopoulos
(F-I) terms, and p-form condensation~\cite{P}.

The generation of a non-perturbative scalar potential, which could stabilize
flat directions of supersymmetric vacua, opens the way to studying
dynamically generated mechanisms of supersymmetry breaking, and to further
explore the possibility of breaking N=2 supersymmetry to N=1.

In a previous note we focused our attention on a minimal model in which the
simultaneous occurrence of the Higgs and super-Higgs mechanisms was shown to
be a necessary condition for triggering the breaking of N=2 to N=1, and for
the lifiting of two of the original six flat directions of the theory.

The model considered there was a spontaneously broken phase of N=2
supergravity coupled to a $U(1)$ vector multiplet and a hypermultiplet,
charged with respect to the $U(1)^2$ gauge group of the theory.
Prior to gauging, the moduli space of the model was the six-dimensional
manifold
\beq
{SO(4,1)\over SO(4)}\times {SU(1,1)\over U(1)}.
\eeq{1}
After gauging, the theory consisted of an N=1 supergravity theory , with a
massive spin-3/2 multiplet together with two massless chiral multiplets,
whose scalars provided the residual (two complex) flat directions of the
theory.
In N=2 supergravity, the presence of a charged hypermultiplet is necessary
to Higgs the $U(1)^2$. This Higgsing is needed to give a mass to
the two spin-1 components of the spin-3/2 massive multiplet~\cite{FV}.

On the other hand, very recently, it was observed in ref.~\cite{APT} that
spontaneous breaking of N=2 rigid supersymmetry to N=1 can already occur in
a self-interacting (non-renormalizable) $U(1)$ abelian theory, if
F-I terms with simultaneous electric and magnetic components are introduced.

In the minimal theory of such kind, the photino is the goldstino and the
broken phase consists of just a massless spin (1,1/2) multiplet together
with a massive chiral multiplet.

This theory is in apparent violation of  ``common wisdom,'' asserting that
it is impossible to break spontaneously N=2 to N=1 in rigid theories.
However, Hughes and Polchinski pointed out in ref.~\cite{HP} that a
generalization of the supersymmetry current algebra can contradict this
wisdom. In fact, two- and four-dimensional counterexamples where provided
within string~\cite{HP} and membrane~\cite{HLP} theory.
Aim of the present paper is, first, to re-analyze more carefully
the general principles
of supersymmetry and to show that, in agreement with ref.~\cite{HP}, it
is indeed possible to have partial breaking
of N=2 in rigid theories; then, to show that such possibility is in fact
encompassed by the previously studied partial breaking in N=2
supergravity~\cite{FGP,CGPe},
if a suitable flat limit, which preserves N=2 rigid supersymmetry, is taken.

Section 1 will re-examine rigid extended supersymmetry, while Section 2 will
deal with local N=2 supergravity and its flat limit.
\section{Partial SUSY Breaking in the Rigid Theory}
Let us recall the argument that forbids partial braking in rigid
supersymmetry. It is a simple consequence of the current
algebra of extended supersymmetry, that implies, among other things, the
following equation:
\beq
\int d^3 \vec{y}\{J_{B\,0\dot{\alpha}}(y^0,\vec{y}), J_{\mu \alpha}^A(x)\}=
2\sigma^\nu_{\alpha
\dot{\alpha}} \delta^A_B T_{\mu\nu}(x),\;\;\; A,B=1,...,{\rm N}.
\eeq{2}
This equation means that the variation under the $B$-supersymmetry of the
$A$-th supercurrent is diagonal in the extension indices, and equal to the
gamma-trace of the stress-energy tensor. This equation, which makes sense
even when supersymmetry is broken, due to the fact that commutators of
fields are local, immediately implies that if a supersymmetry is broken,
then all of them are broken with the same strength. To see this, one takes
the VEV of eq.~(\ref{2}) and inserts a complete basis of states in the
commutator. Standard manupulations imply the existence of N spin-1/2
zero-mass particles (the goldstini), which all couple to the vacuum with equal
strength, $F_{goldstino}^2 \sim \langle 0| T^\mu_\mu |0\rangle$.

The way out to this situation
is that eq.~(\ref{2}) is not the most general current algebra
consistent with supersymmetry~\cite{HP}. Indeed, the Jacobi identities of
supersymmetry~\cite{HLS}
allow for an additional field-independent, constant term
to be  added to eq.~(\ref{2}): $\sigma_{\mu\,\dot{\alpha}\alpha} C_A^B$.
This term does not modify the supersymmetry algebra {\em on the
fields}~\cite{L},
since its commutators with any quantity in the theory is obviously zero.
In particular, the commutator of two supersymmetry transformations on
any field {\em is} a translation (up to eventual gauge transformations).

The presence of this extra constant term in the supersymmetric current
algebra~\cite{FZ}
allows for the breaking of only some of the N supersymmetries.
As we are going to see in a moment, the model of ref.~\cite{APT} (APT model)
precisely realizes this situation. The N=2 supersymmetry is realized
{\em manifestly} and {\em linearly} on the fields
(indeed, the model can be written in terms of N=2 superfields~\cite{APT}),
but the current algebra of the N=2 supersymmetry is not the standard one,
but the one with the additional constant term.

The APT model consists of one (or more) N=2 vector multiplets, $A^\Sigma$.
They can be
written as constrained N=2 chiral multiplets, obeying the constraint:
\beq
(\epsilon_{ij}D^i\sigma_{\mu\nu}D^j)^2 A^\Sigma=-96 \Box A^{*\Sigma}.
\eeq{3}
The crucial observation in ref.~\cite{APT} is that this constraint does not
imply that the auxiliary fields of the vector multiplet,
$\vec{Y}^\Sigma$ -- which transform as vectors of the extension algebra
$SU(2)$ -- are real. A {\em constant} imaginary term is also allowed, so that
$\vec{Y}^{*\Sigma} \neq \vec{Y}^\Sigma$.
This term is a magnetic F-I term, while the real
part of $\vec{Y}^\Sigma$ is a standard N=2 electric F-I term.
The transformation law of the ``gaugino'' (i.e. the spin-1/2 field of the
vector multiplet) is
\beq
\delta \lambda_A^\Sigma= {i\over \sqrt{2}}X_{AB}^\Sigma\eta^B\equiv
{i\over \sqrt{2}}Y_I^\Sigma\epsilon_{AC}\sigma^{I\,C}_B
\eta^B+....\, .
\eeq{4}
Here $I=1,2,3$ and the supersymmetry parameter is $\eta^A$, $A=1,2$.
The ellipsis denote terms which vanish on translationally invariant
backgrounds.
Since $X_{AC}^\Sigma X_{CB}^\Sigma=\vec{Y}^\Sigma\cdot \vec{Y}^\Sigma
\delta_{AB}$ (no sum on $\Sigma$), we see that when
$\vec{Y}^\Sigma$ is real, one unbroken supersymmetry implies
$\vec{Y}^\Sigma=0$, i.e. that the two supersymmetries are both unbroken.
On the other hand, a complex $\vec{Y}$~\footnotemark
\footnotetext{Notice that the off-shell algebra still closes, since an
imaginary constant does not contribute to $\delta \vec{Y}^\Sigma$.}
may have square equal to zero without
vanishing. In ref.~\cite{APT}, in particular, there is a single vector
multiplet, and at the minimum $\vec{Y}=2m\Lambda^2(0, i, -1)$, $m\neq 0$.
Here we have explicitly shown the dependence of $\vec{Y}$ on the
supersymmetry breaking scale $\Lambda \sim F_{goldstino}^{1/2}$.
With this choice of $\vec{Y}$, the matrix $X_{AB}$ has exactly one zero
eigenvalue, i.e. N=2 is broken to N=1.

The lagrangian of the APT model, extended to contain an arbitrary number of
vector multiplets, reads, in N=2 superfield notation~\cite{GSW}:
\beq
{\cal L}={i\over 4} \int d^2\theta_1 d^2\theta_2[{\cal F}(A^\Sigma)
-A_{\Sigma}^D A^\Sigma] + {1\over 2} (\vec{E}_\Sigma\cdot \vec{Y}^\Sigma+
\vec{M}^\Sigma\cdot \vec{Y}_\Sigma^D) + c.c.\, ,
\eeq{5}
where $A^\Sigma$ is an {\em unconstrained} chiral N=2 multiplet,
$A_\Sigma^D$ is a constrained chiral multiplet playing the role of
Lagrange multiplier, $\vec{E}$, $\vec{M}$ are {\em constant} vectors,
and $\vec{M}^\Sigma$ is real.
The equations of motion for the auxiliary fields derived from
lagrangian~(\ref{5}) read:
\bea
\vec{Y}^\Sigma &=& -2\tau_2^{\Sigma\Delta}(\Re \vec{E}_\Delta + \tau_{1\,
\Delta \Gamma}\vec{M}^\Gamma) + 2i\vec{M}^\Sigma, \nonumber \\
\tau_{1\, \sigma \Delta}&=&\Re\tau_{\Sigma\Delta},\;\;\;
\tau_{2\, \Sigma \Delta}= \Im \tau_{\Sigma\Delta}, \;\;\;
\tau_{\Sigma\Delta}={\cal F}_{\Sigma \Delta}, \;\;\;
\tau_2^{\Sigma\Gamma}\tau_{2\, \Gamma\Delta}=\delta^\Sigma_\Delta.
\eea{6}
In computing the supersymmetric variation of the supercurrent, the only
difference with respect to the standard case ($\vec{M}^\Sigma=0$) arises in
terms involving $\vec{Y}^\Sigma$. In other words, all fermionic and
derivative terms in the variation of the supercurrent are as in the standard
case, so that the variation of the supercurrent reads
\beq
\int d^3 \vec{y}\{J_{B\, 0\dot{\alpha}}(y^0,\vec{y}), J_{\mu
\alpha}^A(x)\}=
{1\over 2}\sigma_{\mu\, \alpha\dot{\alpha}}(X^\Sigma_{AC})^{*}\tau_{2\,
\Sigma\Delta}X_{CB}^\Delta + ....=2\sigma_{\mu\,
\alpha\dot{\alpha}}M^A_B+....,
\eeq{7}
where $X_{AB}^\Sigma$ is as in eq.~(\ref{4}), and the ellipsis
denote terms identical with the standard case.
Using eqs.~(\ref{4},\ref{6}) one finds
\bea
M^A_B&=& {1\over 4}\tau_{2\,\Sigma\Delta} \vec{Y}^{*\Sigma}\cdot
\vec{Y}^\Delta\delta^A_B
             + {i\over 4}
\tau_{2\, \Sigma\Delta}\vec{\sigma}^A_B\cdot (\vec{Y}^{*\Sigma}
\times \vec{Y}^\Delta)= \nonumber \\ &=&
\tau_2^{\Sigma\Delta}(\Re \vec{E}_\Sigma +
\tau_{\Sigma\Gamma}\vec{M}^\Gamma)(\Re \vec{E}_\Delta +
\tau^*_{\Delta\Pi}\vec{M}^\Pi) \delta^A_B +2 \vec{\sigma}^A_B\cdot ( \Re
\vec{E}_\Sigma \times \vec{M}^\Sigma) .
\eea{8}
The term proportional to $\delta^A_B$ is the scalar potential, as in the
standard case.
Since $M^A_B$ is given by the square of the fermionic shifts in
eq.~(\ref{4}), it is positive semidefinite.
Its eigenvalues are
\beq
\lambda_\pm=
{1\over 4}\tau_{2\, \Sigma\Delta}\vec{Y}^\Sigma \cdot \vec{Y}^{*\Delta} \pm
{1\over 4}
|| i\tau_{2\, \Sigma\Delta} \vec{Y}^\Sigma \times \vec{Y}^{*\Delta} ||
, \;\;\; ||\vec{V}||\equiv (\vec{V}\cdot \vec{V})^{1/2}.
\eeq{8a}
Partial breaking is possible whenever $\lambda_-=0$.

By substituting eq.~(\ref{8}) into the variation of the
supercurrent we find that the supersymmetry current algebra receives a
{\em field-independent} modification, as expected:
\beq
\int d^3 \vec{y}\{J_{B\, 0\dot{\alpha}}(y^0,\vec{y}), J_{\mu \alpha}^A(x)\}=
2\sigma^\nu_{\alpha
\dot{\alpha}} \delta_B^A T_{\mu\nu}(x) + 4\sigma_{\mu\, \alpha\dot{\alpha}}
\vec{\sigma}^A_B\cdot ( \Re\vec{E}_\Sigma \times \vec{M}^\Sigma) .
\eeq{9}
As we noticed before, this additional constant term in the current algebra
does not affect the form of the supersymmetry transforamtions on fields.
Indeed, in the presence of a nonzero $\vec{M}$, the commutator of two
supersymmetry transformations still has the standard form on all fields
except on the spin-1 field, where it gets an extra term
\beq
[\delta_1, \delta_2]A^\Sigma_\mu=
2i (\eta^1_A\sigma_\mu\bar{\eta}^{2B}-\eta^2_A\sigma_\mu \bar{\eta}^{1B})
\vec{\sigma}_B^A \cdot \vec{M}^\Sigma.
\eeq{10}
This extra term is a harmless gauge transformation when the supersymmetry
parameters are constant. This is no longer the case when they are given a
dependence on space-time coordinate; in other words, the APT model cannot
be coupled naively to supergravity. We shall see in a moment that in spite
of this, it can be recovered as an appropriate flat limit of N=2
supergravity.
\section{N=2 Supergravity with Partial Breaking and its Flat Limit}
In this section we show that the APT model arises as a flat limit of a N=2
supergravity. The limit is $M_{Pl}\rightarrow \infty$,
$\Lambda=\mbox{constant}$, where, as before, $\Lambda\sim F_{goldstino}^{1/2}$
is the scale of supersymmetry breaking.
For sake of simplicity we will study the original
APT theory, which contains a single self-interacting abelian vector multiplet.

As we have just seen, the current algebra (i.e. the Ward identity)
of rigid supersymmetry can be modified by adding a constant term to it.
In local supersymmetry, this
freedom no longer exists, instead, the algebra is modified because of the
presence of the gravitini. When restricted to translationally invariant
backgrounds, the Ward identity becomes to so-called
``T-identity''~\cite{CGP,FM}
\beq
\delta_A\psi_L^i\delta^B\psi_R^j{\cal Z}_{ij}-3M_{Pl}^2
{\cal M}_{AC}{\cal M}^{*\, CB}=V\delta_A^B,
\eeq{11}
where $\delta_A\psi^i$ denote the shift, under the $A$-th
supersymmetry, of the spin
one-half fermions, while ${\cal M}_{AB}$ is the gravitino mass matrix and
${\cal Z}_{ij}$ is the kinetic term of the fermions. We have kept the
dependence on $M_{Pl}$ in this formula because we will be interested in
studying an appropriate flat limit $M_{Pl}\rightarrow \infty$.
These identities show that even when $V=0$, one may still have, say
\beq
\delta_1\psi^i_L\delta^1\psi^j_R{\cal Z}_{ij}=3{\cal M}_{1C}{\cal M}^{C1}=0,
\eeq{12}
but instead
\beq
\delta_2\psi^i_L\delta^2\psi^j_R{\cal Z}_{ij}=3{\cal M}_{2C}{\cal M}^{C2}\neq
0.
\eeq{13}
In N=2, this corresponds to breaking half of the supersymmetries
(N=1 unbroken), at zero cosmological constant.

We must emphaisze that a field configuration for which at least one
supersymmetry is unbroken gives automatically an absolutely stable local
minimum of the potential~\cite{CGP}.

A supergravity model with partial breaking of N=2 supersymmetry can be
constructed along the lines of ref.~\cite{FGP}.
The matter content of the model is a charged hypermultiplet, whose
scalars parametrize the quaternionic manifold $SO(4,1)/SO(4)$, coupled to an
abelian vector multiplet.

By denoting the quaternionic coordinates of the hypermultiplet manifold
with  $b^u$, $u=0,1,2,3$, we can write its symplectic vielbein~\cite{DFF}
-- which determines the coupling to fermions -- as~\cite{FGP}
\beq
{\cal U}^{\alpha A}={1\over 2b^0} \epsilon^{\alpha\beta}(db^0
-i\vec{\sigma}d\vec{b})_\beta^{\; A}.
\eeq{14}

The special geometry of the vector-multiplet manifold is specified by
four holomorphic sections $X^\Sigma(z)$, $F_\sigma(z)$,
$\Sigma=0,1$~\cite{CDFV}, in terms of which the K\"ahler potential reads
\beq
K=-\log i({X}^{*\Lambda} F_\Lambda   -X^\Lambda {F}^*_\Lambda).
\eeq{15}
Our choice, which slightly generalizes that of ref.~\cite{FGP}, is
\beq
X^0(z)={1\over \sqrt{2}},\;\;\;  X^1(z)={i\over \sqrt{2}} f'(z),\;\;\;
F_0(z)=-{i\over \sqrt{2}}[2f(z)-zf'(z)],\;\;\; F_1(z)={z\over \sqrt{2}}.
\eeq{16}
This choice of sections is such that no holomorphic prepotential exists. It
has been obtained by performing the symplectic transformation
$X^1\rightarrow -F_1$, $F_1\rightarrow X^1$ on the sections obtained from
the prepotential $F(X^0,X^1)=-i(X^0)^2f(X^1/X^0)$~\cite{FGP,CDFV}.

The gauge group in our case is $U(1)^2$, with one of the $U(1)$ factors
coming from the matter vector multiplet and the other from the graviphoton.
The coupling of $U(1)^2$ to the hypermultiplet is specified by the covariant
derivative $D_\mu b^u=\partial_\mu b^u + A_\mu^\Sigma k^u_\Sigma$. The
Killing vectors $k^u_\Sigma$ in our case are a simple generalization of
those of ref.~\cite{FGP}:
\beq
k^u_0=g_1\delta^{u3}+g_2\delta^{u2},\;\;\; k^u_1=g_3\delta^{u2}.
\eeq{17}
$g_1$, $g_2$ and $g_3$ are arbitrary constants.
These Killing vectors are derived through standard N=2 formulae from the
``D-term'' prepotentials~\cite{DFF,FGP}
\beq
{\cal P}^I_0={1\over b^0}(g_1\delta^{I3}+g_2\delta^{I2}),\;\;\;
{\cal P}^I_1=g_3{1\over b^0}\delta^{x2},\;\;\; I=1,2,3,
\eeq{18}
i.e. $\vec{\cal P}_\Sigma=b_0^{-1}\vec{k}_\Sigma $.

The formulae for the shifts of the (antichiral) gaugino $\lambda^{z^*}_A$,
hyperini $\zeta^\alpha$, and (chiral) gravitini $\psi_{A\, \mu}$ are
\bea
\delta \lambda^{z^*}_A &=&
-ig^{zz^*}\epsilon_{BC}\vec{\sigma}_A^C \cdot\vec{\cal
P}_\Sigma e^{K/2}(\partial_z + \partial_z
K)X^\Sigma(z)\eta^B=W^{z^*}_{AB}\eta^B, \nonumber \\
\delta \zeta^\alpha &=& -2 \epsilon_{AB}{\cal U}^{\alpha B}_u k^u_\Sigma
e^{K/2}X^\Sigma(z)\eta^A \equiv {\cal N}_A^\alpha \eta^A,  \nonumber\\
\delta \psi_{A\, \mu}&=& {i\over 2} \epsilon_{BC}\vec{\sigma}_A^C
\vec{\cal P}_\Sigma e^{K/2}X^\Sigma(z)\gamma_{\mu}\eta^B\equiv
iS_{AB}\gamma_{\mu}\eta^B.
\eea{19}
To recover the APT model, we must expand around an appropriate vacuum, with
zero cosmological constant at the Planck scale, and
perform a suitable flat limit $M_{Pl}\rightarrow \infty$ .
It is evident that the coupling constants
$g_i$, as well as the gravitino mass, must vanish in that limit. The reason
is that the APT model does not contain any hypermultiplet, thus, to
reconcile its spectrum with the one of our supergravity theory, the
hypermultiplet must decouple from the vector multiplet in the flat limit.
Moreover, a finite, nonzero gravitino mass would imply that the lagrangian
contains {\em explicit} supersymmetry-breaking terms in the
$M_{Pl}\rightarrow \infty$ limit. The crucial property that allows for a
non-trivial result in the flat limit is that the gravitino shifts are
proportional to the supersymmetry breaking scale $\Lambda$, while the latter
is related to the gravitino mass by  $m_{3/2}\sim \Lambda^2/M_{Pl}$.
It is therefore possible to find a limit in which the
gravitino shifts contribute to the Ward identity of rigid supersymmetry,
while the gravitino mass goes to zero (together with all explicit
supersymmetry-breaking terms).

The limit reproducing the APT model is specified as follows.
First of all we set
\beq
f(z)= {1\over 2} + {\Lambda\over M_{Pl}} z + {\Lambda^2 \over
M_{Pl}^2} \phi(z) + O(\Lambda^3/M_{Pl}^3).
\eeq{20}
Then we choose
\beq
g_1={\Lambda^2\over M_{Pl}^2} \xi, \;\;\; g_2= {\Lambda^2\over
M_{Pl}^2} e, \;\;\; g_3= 2{\Lambda \over M_{Pl}}m.
\eeq{21}
In the limit $M_{Pl}\rightarrow\infty$, $\Lambda=\mbox{constant}$, we find
\bea
g_{zz^*}&=&\partial_z\partial_{z^*}K(z,z^*)= {\Lambda^2\over
M_{Pl}^2}\left\{ 1-{1\over 2} \phi''(z) -{1\over 2} [\phi''(z)]^* \right\}
+ O(\Lambda^3/ M_{Pl}^3), \nonumber \\
K_{z}&=& -{\Lambda\over M_{Pl}} + O(\Lambda^2/M_{Pl}^2).
\eea{22}
To recover the APT model we set
\beq
{\cal F}(z)\equiv z^2 - i 2\phi(z),
\eeq{23}
and we rescale the fermions so that they have a canonically normalized
kinetic term:
\beq
\lambda^{z^*}_A \rightarrow (M_{Pl}\Lambda^2)^{-1/2}\lambda^{z^*}_A,\;\;\;
\zeta^\alpha \rightarrow M_{Pl}^{-3/2}\zeta^\alpha,\;\;\;
\psi_{A\,\mu} \rightarrow M_{Pl}^{-3/2} \psi_{A\, \mu}.
\eeq{24}
By restoring the proper mass dimension ($-1/2$) in the supersymmetry breaking
parameter, i.e. rescaling $\eta^A\rightarrow M_{Pl}^{1/2}\eta^A$, and by
defining
$\tau(z)=\tau_1(z) +i\tau_2(z)={\cal F}''(z)$, we find
that $g_{zz^*}=(\Lambda^2/M_{Pl}^2)\tau_2(z)/2$, and
that in the flat limit the shifts of the fermions read
\bea
\delta \lambda^{z^*}_A &=& i\Lambda^2{2\over \tau_2(z)}\epsilon_{BC}
\left[{1\over \sqrt{2}b^0}(\xi\sigma^{3C}_A + e\sigma_A^{2C}) +
{1\over \sqrt{2}b^0} m\sigma^{2C}_A\tau(z)\right]\eta^B
+O(\Lambda^3/M_{Pl}),\nonumber \\
\delta \zeta^\alpha &=& i\Lambda^2\epsilon_{BC}{1\over \sqrt{2}b^0}\left[
(\xi\sigma^{3C}_A + e\sigma^{2C}_A) + 2i m\sigma^{2C}_A\right]\eta^B
+ O(\Lambda^3/M_{Pl}),
\nonumber \\
\delta \psi_{A\, \mu} &=& {i\over 2} \Lambda^2\gamma_\mu
\epsilon_{BC}{1\over \sqrt{2}b^0}\left[
(\xi\sigma^{3C}_A + e\sigma^{2C}_A) + 2i m\sigma^{2C}_A\right]\eta^B
+ O(\Lambda^3/M_{Pl}).
\eea{25}
Notice that in the flat limit, $b^0$ becomes a coupling constant,
since the fluctuations in $\partial_\mu b^0$ are $O(M_{Pl}^{-1}$). Therefore
$b^0$ plays a role analogous to the dilaton in the low-energy limit of string
theory. Since the manifold of the hypermultiplet is homogeneous, all values
of $b^0>0$ give the same physics. In the flat limit the only effect of a
change in $b^0$ is to rescale the coupling constants $\xi$, $e$ and $m$.
In particular, at $b^0=1$, we recover the APT model with a single
vector multiplet and
\beq
\Re \vec{E}=\Lambda^2(0,e,\xi),\;\;\; \vec{M}=\Lambda^2 (0,m,0).
\eeq{26}
Eq.~(\ref{25}) shows that in the flat limit the shift of the gaugino is
identical with the one of the APT model, whereas the gravitino and hyperino
shifts are nonzero, {\em field independent constants}. We are thus guaranteed
that the potential of our model agrees with the APT one, up to field
independent terms.
This statement can be easily verified by using the T-identitites
eq.~(\ref{11}). In the normalizations of this section these identities read
\beq
-12(S_{AC})^*S_{CB} + {\tau_2(z)\over 2}(W^{z^*}_{AC})^* W_{CB}^{z^*} +
2({\cal N}^\alpha_A)^* {\cal N}^\alpha_B =\delta^A_B V(z),
\eeq{27}
where we have used the re-scaled fermion shifts throughout.
As expected, the (field independent) terms
that are off-diagonal in the extension indices $A$, $B$, and that
arise from the square of the gaugino shifts, are exactly cancelled by the
combined shifts of the hyperini and gravitini. The same shifts also
contribute an additional {\em field-independent} term to the scalar potential,
equal to $-(1/2)\Lambda^4( \xi^2 + e^2+ 4m^2)$. The complete potential reads
\beq
V(z)= {1\over b^{0\, 2}}
\left[{1\over \tau_2(z)}|\Re \vec{E} + \tau(z)\vec{M}|^2 -{1\over
2}\Lambda^4(\xi^2 + e^2 + 4m^2)\right].
\eeq{28}

Some comments are now in order.
\begin{enumerate}
\item
The potential in eq.~(\ref{27}) differs from the one
in~\cite{APT} or in eq.~(\ref{8}). Since the difference depends only on
$b^0$, it becomes irrelevant in the flat limit, when gravitational
interactions decouple and $b^0$ does not fluctuate.
The stationary point $\tau_1(z)=-e/m$,
$\tau_2(z)=|\xi/m|$ is a stable minimum~\cite{APT} at $M_{Pl}=\infty$.
At this minimum,
\beq
V_{min}={1\over 2b^{0\, 2}}\Lambda^4(4|\xi m| -\xi^2 - e^2 - 4m^2).
\eeq{29}
\item
Since the hyperino and gravitino shifts depend only on $b^0$, for a generic
choice of $\xi$, $e$, $m$, they break both supersymmetries. On the other
hand, since the gaugino shift depends also on the field $z$, whenever the
``prepotential''
${\cal F}(z)$ gives rise to non-renormalizable interactions, one may be able
to find a VEV $z$ such that the gaugino-shift matrix has exactly one zero
eigenvalue. In this case N=2 supersymmetry is broken to N=1 in both the
``observable'' gaugino sector and the ``hidden'' hyperino + gravitino
sector, while the residual N=1 is only broken in the hidden sector.
\item
The cosmological constant is effectively zero at the Planck scale, whenever
$\Lambda \ll M_{Pl}$. Nevertheless, for $M_{Pl}<\infty$,
and for generic values of the parameters $\xi$, $e$,
$m$, the potential in eq.~(\ref{28}) has no minimum. Indeed, in the full
supergravity theory described above, $b^0$ is a dynamical field.
Minimization in $b^0$ and $z$ result in a constraint on the parameters
of the theory, which reads, up to terms $O(\Lambda/M_{Pl})$:
\beq
4|\xi m| -\xi^2 - e^2 - 4m^2=0.
\eeq{30}
This equation is solved by $e=0$, $\xi=2m$. These parameters define a
supergravity model undergoing partial super-Higgs from N=2 to N=1: N=1 is
unbroken in both the ``observable'' sector and the ``hidden'' sector.
A general
theorem~\cite{CGP} guarantees that in this case the stationary point
in $z$ and $b^0$ exists and is a stable minimum.
\item
For any other choice of the parameters, only ``cosmological'' solutions with
a time-dependent runaway $b^0$ VEV exist.
The characteristic time of evolution of these solutions, $db^0/dt$, is
$\kappa M_{Pl}/\Lambda^2$, where $\kappa$ is $O(1)$ for a generic choice of
parameters.
If we set $\Lambda \sim 100\, GeV$, this time is
approximately $10^{-10}\, sec$. Needless to say, this means that a
``realistic'' supersymmetry breaking scale does not give rise to
a realistic -- i.e. almost stationary -- perturbative vacuum.
\item
The model of ref.~\cite{FGP} has an accidental flat direction, giving rise
to an extra massless N=1 scalar multiplet, due to a special choice of the
vector-multiplet metric. In ref.~\cite{FGP}, indeed, the vector multiplet
manifold was the homogeneous space $SU(1,1)/U(1)$, corresponding to choosing
$f(z)=(1/2) + (\Lambda/M_{Pl})z$ in eq.~(\ref{20}).
\end{enumerate}

\vskip .2in
\noindent
Acknowledgements
\vskip .1in
\noindent
We would like to thank C. Imbimbo and A. Zaffaroni for useful discussions and
comments.

S.F. is supported in part by DOE under grant DE-FGO3-91ER40662, Task C, and by
EEC Science Program SC1*-CI92-0789. L.G. is supported in part by
Ministero dell' Universit\`a e della Ricerca Scientifica e Tecnologica,
by INFN, and by EEC Science Programs SC1*-CI92-0789 and CHRX-CT92-0035.
M.P. is supported in part by NSF under grant PHY-9318781.

\end{document}